\documentclass[fleqn,epsf]{article}
\newfont{\mainfont}{cmss10 scaled 1000}      
\newfont{\titlefont}{cmss12 scaled 1600}   
\newfont{\authorfont}{cmcsc10 scaled 1000}   
\newfont{\authoraddfont}{eurm10 scaled 1000} 
\newfont{\Labelfont}{cmss12 scaled 1200}   
\newfont{\labelfont}{cmss12 scaled 1100}   
\newfont{\figfont}{cmss10 scaled 1000}       
\newfont{\footfont}{cmss10 scaled 1000}      
\newfont{\tabfont}{cmss10 scaled 1000}       
\newfont{\reffont}{cmss10 scaled 1000}       
\newfont{\refnamefont}{cmss12 scaled 1100} 
\newfont{\abstractfont}{cmss12 scaled 1000}  
\newfont{\abstractnamefont}{eurm10 scaled 1200} 
    \newcommand{\ba}{\begin{eqnarray}}
    \newcommand{\ea}{\end{eqnarray}}
    \newcommand{\be}{\begin{equation}}
    \newcommand{\ee}{\end{equation}}

    \newcommand{\psb}{\bar{\psi}}
    
\newcommand{\AmS}{{\protect\the\textfont2
  A\kern-.1667em\lower.5ex\hbox{M}\kern-.125emS}}

\title{ \titlefont Hybrid Monte Carlo algorithm for lattice QCD with two flavors 
of dynamical Ginsparg-Wilson quarks}
\author{
            {\authorfont Chuan Liu}\\ 
  {\authoraddfont Department of Physics}\\
  {\authoraddfont Peking University}\\
  {\authoraddfont Beijing, 100871, P{\mainfont .}~R{\mainfont .}~China}
         }
       
\begin{document}

\maketitle

\begin{abstract}
\setlength{\baselineskip}{0.65cm}

\abstractfont{
We study aspects concerning numerical simulations of 
Lattice QCD with two flavors of
dynamical Ginsparg-Wilson quarks with degenerate masses.
A Hybrid Monte Carlo algorithm is described and the formula for
the fermionic force is derived for two specific implementations.
The implementation with optimal rational approximation method
is favored both in CPU time and memory consumption.
} 
\end{abstract} 

\newpage


\section{Introduction}

Recently, in a series
of publications\cite{neuberger,hasenfratz,hasenfratz2,hasenfratz3,
luscher,chandra}, 
it has become clear that, if one would
modify the chiral transformation laws away from their canonical form in
the continuum, chiral symmetries can be preserved on the lattice without
the problems of fermion doubling.
The lattice Dirac operator $D$ for 
these fermions satisfies the Ginsparg-Wilson \cite{ginsparg}
relation\footnote{They are therefore named Ginsparg-Wilson fermions.}, 
\be
\label{eq:ginsparg}
\gamma_5 D +D\gamma_5 D =aD \gamma_5 D  \;\;,
\ee
where $a$ is the lattice spacing. 
As a consequence of the  
Ginsparg-Wilson relation~(\ref{eq:ginsparg}), it is easy 
to show that the fermion action,
\be
S_F=\sum_{x,y}\psb(x) D_{x,y} \psi(x) \;\;,
\ee
is invariant under lattice chiral transformations, and 
chiral symmetry will protect the quark masses away from the
additive renormalizations.

The chiral properties of the Ginsparg-Wilson fermions is
a direct result of the Ginsparg-Wilson relation. In particular, several
types of fermion actions could be written down which all fulfill the
condition. For definiteness, one particular choice \cite{neuberger2,luscher} 
is adopted in this paper, namely: 
\ba
\label{eq:ferm_matrix}
aD&=&1-H_W(H^{\dagger}_W H_W)^{-1/2} \;\;, \nonumber \\
H_W&=& 1+s-a\sum_{\mu}{1 \over 2}
[\gamma_\mu(\nabla_\mu+\nabla^*_\mu)-a\nabla^*_\mu\nabla_\mu] \;\;,
 \nonumber \\
&=&(-3+s)\delta_{x,y}+
{1 \over 2}\sum_{\mu}\left[(1-\gamma_\mu)U_{\mu}(x)\delta_{x+\mu,y}
+(1+\gamma_\mu)U^{\dagger}_{\mu}(x-\mu)\delta_{x-\mu,y}\right]
\ea
where $s$ is a parameter satisfying $|s|<1$.  
The lattice covariant derivatives
$\nabla^*_\mu$ and $\nabla_\mu$ are defined as usual,
\ba
\nabla_\mu \psi&=& U_\mu(x)\psi(x+\mu)-\psi(x) \;\;, \nonumber \\
\nabla^*_\mu \psi&=& \psi(x)-U^{\dagger}_{\mu}(x-\mu)\psi(x) \;\;. 
\ea

Due to their decent chiral properties, it is quite tempting to investigate
the possibility of performing Monte Carlo simulations using 
Ginsparg-Wilson fermions, replacing the conventional
Wilson fermions.  Albeit their
seemingly non-local appearance, the fermion matrix~(\ref{eq:ferm_matrix})
is in fact local (with exponentially decaying tails), 
as long as the parameter $s$
in the matrix is chosen appropriately
\cite{luscher2,chiu}. The locality property of the
 fermion matrix will enable us to use iterative methods in
Krylov space whenever inversion of the matrix becomes necessary.

When performing the inversion of the fermion matrix~(\ref{eq:ferm_matrix}), 
one would encounter the problem of yet another matrix 
inversion of a fractional power.
Recently, proposals have been put forward \cite{neuberger2,bunk} which
make  such inversions possible. 
Some numerical calculations in quenched QCD \cite{edwards}
for Ginsparg-Wilson fermions already indicate that these fermions
indeed  have anticipated chiral properties.
However, it is also realized that quenched calculations with Ginsparg-Wilson
fermions are more costly than with conventional Wilson fermions
primarily due to fractional inversion of 
the matrix $H^{\dagger}_WH_W$.
Two specific types of methods will be discussed in this paper. One is the
method proposed by Bunk \cite{bunk}, which will be called 
fractional inverter method, or FIM. The other method is the
optimal rational approximation method, or ORAM,
proposed in Ref.\cite{edwards}.
It was reported in Ref.~\cite{edwards} that ORAM converges faster
than FIM for a desired accuracy and a given condition number of
the matrix $H^{\dagger}_WH_W$.
Now, each multiplication with the fermion matrix $D$
for Ginsparg-Wilson fermions is
equivalent to $2N+1$ multiplications with matrix $H^{\dagger}_W$ or
$H_W$, where $N$ is some integer.
With FIM, $N$ is the number of 
highest order Legendre polynomial kept in the iteration procedure.
With ORAM, $N=N_{cg}$ is the number of conjugate gradient \footnote{
By the phrase "conjugate gradient", we mean all possible iterative
algorithms in Krylov space: conjugate gradient, minimal residue,
Bi-conjugate gradient, etc.} 
iterations needed to perform the multi-shift matrix inversions.
This number is determined by
the condition number of the matrix  $H^{\dagger}_WH_W$ and
the accuracy desired.
Therefore, the calculations with Ginsparg-Wilson fermions is
at least more costly than conventional Wilson fermions
by a factor of $2N+1$.

Although it is already quite costly in the quenched case, it remains 
an tempting problem to simulate {\em dynamical} 
Ginsparg-Wilson fermions.  No algorithms including dynamical fermions
 have been tested on these newly proposed fermions.
In this paper, it is shown that a Hybrid Monte Carlo algorithm would do
the job, however, just as in the quenched case, the simulation is
more costly than simulating {\em dynamical} Wilson fermions.
Also, in the calculation of 
the fermionic force to gauge links, using
two different methods for the fractional inversion
results in very different memory and CPU time consumptions.
For FIM, it seems that
 $O(N)$ psudofermion fields have
to be stored in order to make the simulation tractable.
For ORAM, only a moderate number of
psudofermion fields have to be stored, and the number of matrix
multiplications also increases slower than in FIM.
In this paper, we will concentrate on a Hybrid Monte Carlo
algorithm for simulation of dynamical Ginsparg-Wilson fermions.
The fermionic force for the gauge links, which is the crucial
part for the dynamical fermion simulation, will be calculated 
for both ORAM and FIM. General properties of the two methods 
are compared. Test runs on small lattices with gauge group
$SU(3)$ are now being investigated \cite{chuan2} and detailed
results will be reported later.

This paper is organized in the following manner. In Section 2, 
the Hybrid Monte Carlo algorithm suitable for 
simulating dynamical Ginsparg-Wilson fermions
are described and the formula for the force of the gauge link is 
derived, in both ORAM and FIM. These two methods are compared
in the dynamical simulation in terms of CPU time consumption
and memory consumption. Possible improvement methods are
also addressed. Some concluding remarks are in Section 3.

\section{ The Hybrid Monte Carlo algorithm}

The basic formalism of Hybrid Monte Carlo
algorithm \cite{kennedy} remains the same
as in the conventional Wilson case \cite{chuan}.
Only the fermionic force has to be re-derived for the
Ginsparg-Wilson case, which will be dealt with below in detail.
The effective action with the psudofermion contribution now reads:
\be
S_{eff}=S_g[U_{\mu}(x)]+\phi^{\dagger}Q^{-2}\phi \;\;,
\ee
where the fermion matrix $Q \equiv \gamma_5(D+m)$ is hermitian and
$\phi(x)$ being the psudofermion field generated at the beginning of
a Hybrid Monte Carlo trajectory from  Gaussian noise.
We have also assumed that two flavors of quarks have degenerate masses. 
At each molecular dynamics step in a Hybrid Monte Carlo trajectory,
one has to find solution vectors 
$X_1 \equiv Q^{-2}\phi$ and  $X_2 \equiv Q^{-1}\phi=QX_1$ from 
an iterative algorithm (conjugate gradient, for example) in Krylov space.
Then, the total force with respect to the gauge fields can be found
by investigating the variation of the action under infinitesimal 
changes of the gauge fields:
\ba
\delta S_{eff}&=&\sum_{x,\mu} Tr[V_{\mu}(x)\delta U_{\mu}(x)+h.c.]
+  \delta S_f \;\;, 
\nonumber \\
\delta S_f&=& \delta[\phi^{\dagger}Q^{-2}\phi]
=  Tr[F_{\mu}(x)\delta U_{\mu}(x)+h.c.] \;\;. 
\ea
The gauge staple $V_{\mu}(x)$ comes solely from
the pure gauge part of the action and could be
obtained with little cost. The fermionic forces $F_{\mu}(x)$, 
however, is much more costly. Once the fermionic forces
are obtained, the standard Hybrid Monte Carlo updating procedure
can be carried on just as in the conventional Wilson case.

To derive the formula for the fermionic force, we take the variation
of the fermionic part of the action and get,
\be
\delta S_f= X^{\dagger}_1(-\delta Q) X_2 + X^{\dagger}_2(-\delta Q) X_1 
\ee
The variation of the matrix $Q$ contains two parts, one being 
simple, namely 
\be
-\delta_1 Q = \gamma_5 (\delta H_W) (H^{\dagger}_WH_W)^{-1/2} \;\;,
\ee
the other being quite complicated, i.e.
\be
-\delta_2 Q = \gamma_5 H_W \delta (H^{\dagger}_WH_W)^{-1/2} \;\;,
\ee
which depends on the detailed implementation of the fractional
inversion of the matrix. We now proceed to discuss the fermionic
forces in ORAM and FIM respectively.

\subsection{Fermionic force in ORAM}

We first present the force in ORAM which is more straightforward.
We recall that, this approximation 
amounts to approximate the function  $z(z^2)^{-1/2}$ in the
interval $[0,1]$ by a ratio of two polynomials:
\be
z(z^2)^{-1/2} = z\left(c_0+\sum^N_{k=1}{c_k \over z^2+q_k}\right) \;\;,
\ee
It is an approximation similar to the Pad\'e approximation
used in Ref.\cite{ying}. For details about this approximation
and how to get coefficients $c_k$, consult~\cite{edwards}
and references therein.
Applying this method to the hermitian matrix $\gamma_5H_W$,
we immediately obtain the following expression for
the variation of the fermionic action:
\ba
\delta S_f &=& c_0Tr(X_2 \otimes X^{\dagger}_1\gamma_5\delta H_W)
+\sum^N_{k=1}c_k Tr(\zeta_{2,k} \otimes X^{\dagger}_1 \gamma_5 \delta H_W)
\nonumber \\
&+&c_0Tr(X_1 \otimes X^{\dagger}_2\gamma_5\delta H_W)
+\sum^N_{k=1}c_k Tr(\zeta_{1,k} \otimes X^{\dagger}_2 \gamma_5 \delta H_W)
\nonumber \\
&-&\sum^N_{k=1}c_k Tr\left( \zeta_{2,k} \otimes \xi^{\dagger}_{1,k} H^{\dagger}_W \delta H_W
+ \xi_{2,k} \otimes \xi^{\dagger}_{1,k} \gamma_5 \delta H_W
\right) \;\;,
\nonumber \\
&-&\sum^N_{k=1}c_k Tr\left( \zeta_{1,k} \otimes \xi^{\dagger}_{2,k} H^{\dagger}_W \delta H_W
+ \xi_{1,k} \otimes \xi^{\dagger}_{2,k} \gamma_5 \delta H_W
\right) \;\;,
\nonumber \\
\zeta_{i,k}&=& {1 \over (\gamma_5H_W)^2+q_k}X_i \;\;, \;\;\;\;
\xi_{i,k}= \gamma_5H_W {1 \over (\gamma_5H_W)^2+q_k}X_i \;\;.
\ea
In the above formula, ``$Tr$'' stands for taking trace in both Dirac and color
space and a summation over the whole lattice points.
The symbol $\otimes$ stands for direct product of two vectors in
color space.  Therefore, the fermionic force is obtained as
\ba
F_{\mu}(x)&=&{c_0 \over 2} tr_{Dirac}[X_2(x+\mu) \otimes X^{\dagger}_1(x)\gamma_5(1-\gamma_{\mu}) 
+X_1(x+\mu) \otimes X^{\dagger}_2(x)\gamma_5(1-\gamma_{\mu})]
\nonumber \\
&+&\sum^N_{k=1}tr_{Dirac}{c_k \over 2}[\zeta_{2,k}(x+\mu) \otimes X^{\dagger}_1(x)
\gamma_5(1-\gamma_{\mu})]
\nonumber \\
&+&\sum^N_{k=1}tr_{Dirac}{c_k \over 2}[\zeta_{1,k}(x+\mu) \otimes X^{\dagger}_2(x)
\gamma_5(1-\gamma_{\mu})]
\nonumber \\
&-&\sum^N_{k=1}tr_{Dirac}{c_k \over 2}[\zeta_{2,k}(x+\mu) \otimes 
[H_W\xi_{1,k}]^{\dagger}(x)(1-\gamma_{\mu})]
\nonumber \\
&-&\sum^N_{k=1}tr_{Dirac}{c_k \over 2}[\xi_{2,k}(x+\mu) \otimes \xi^{\dagger}_{1,k}(x)
\gamma_5(1-\gamma_{\mu})]
\nonumber \\
&-&\sum^N_{k=1}tr_{Dirac}{c_k \over 2}[\zeta_{1,k}(x+\mu) \otimes 
[H_W\xi_{2,k}]^{\dagger}(x)(1-\gamma_{\mu})]
\nonumber \\
&-&\sum^N_{k=1}tr_{Dirac}{c_k \over 2}[\xi_{1,k}(x+\mu) \otimes \xi^{\dagger}_{2,k}(x)
\gamma_5(1-\gamma_{\mu})] \;\;,
\ea
where the trace $tr_{Dirac}$ is only taken within the Dirac space.

\subsection{Fermionic force in FIM}

In FIM, we would like to solve for $\xi$ satisfying:
\be
\label{eq:solve}
M^{1/2}\xi=X \;\;\;\;, {\rm given} \;\;\; X.
\ee
where the matrix $M \equiv H^{\dagger}_W H_W$ by setting:
\be
M = c (1+t^2-2tA) \;\;,
\ee
with the parameters $c$ and $t$ chosen in such a way 
that all eigenvalues of the matrix $A$ lie within $[-1,1]$. 
To be more specific, we choose,
\be
t = {\sqrt{\kappa}-1 \over\sqrt{\kappa}+1}  \;\;,\;\;\;\;
c = {(\sqrt{\kappa}+1)^2 \over 4 \lambda_{min}} \;\;,
\ee
where $\lambda_{min}$ ($\lambda_{max}$) is the lowest
(highest) eigenvalue of the matrix $H^{\dagger}_WH_W$ and
 $\kappa \equiv \lambda_{max}/\lambda_{min}$ is the condition number.
Then, the solution to eq.~(\ref{eq:solve}) may be written as:
\be
\xi=c^{-1/2} \sum^{\infty}_{k=0} t^k P_k(A) \cdot X
=\sum^{\infty}_{k=0} s_k \;\;,
\ee
where $P_k(z)$ are Legendre polynomials. Therefore, an approximant for the
solution at the $n$-th level is
\be
\xi_n =\sum^{n}_{k=0} s_k \;\;.
\ee
The shifts \footnote{A extra factor $t^k$ has been included
in the definition of $s_k$ as compared with Ref.\cite{bunk}}, 
$s_k$, defined as
\be
s_k =c^{-1/2}t^kP_k(A)\cdot X \;\;,
\ee
satisfy the following recursion relations:
\ba
s_{-1}&=&0 \;\;, s_0=c^{-1/2}X \;\;, \nonumber \\
s_{k}&=&(2-1/k)tAs_{k-1}-(1-1/k)t^2s_{k-2} \;\;.
\ea
For the case of Legendre polynomials, it is claimed that the following
bound for the residue is obtained \cite{bunk}:
\be
|\xi-\xi_n|/|\xi|=|R_n(A)| \le t^{n+1}=
 ({\sqrt{\kappa}-1 \over\sqrt{\kappa}+1})^{n+1} \;\;,
\ee
which asserts the exponential convergence of the iteration.

For the vectors  $\delta (H^{\dagger}_WH_W)^{-1/2}X_i$,
similar strategy could be applied, 
\be
\delta(M^{-1/2})\eta = c^{-1/2}
\sum^{N_{cut}}_{n=0} t^n \delta P_n(A)\eta \;\;, \;\;\;\;\; {\rm given} \;\;\; \eta,
\ee
where $\eta$ represents either $X_1$ or $X_2$ and $N_{cut}$ is
the highest order of Legendre polynomials kept in the approximation.
In an analogous manner, we define,
\be
\delta_k \equiv c^{-1/2} t^k (\delta P_k(A)) \cdot \eta \;\;,
\ee
which satisfy the following recursion relation:
\ba
&&(k+1) \delta_{k+1}+kt^2 \delta_{k-1}=(2k+1)t(\delta As_k+A\delta_k) \;\;.
\nonumber \\
&& \delta_{-1}=0 \;\;\;\;, \delta_0=0\;\;,
\ea
Using this relation, $\delta_k$ could be expressed as:
\be
\delta_k = \sum^{k-1}_{l=0} tL^l_k(A)\delta A s_l \;\;.
\ee
The coefficients $L^l_k(A)$ are polynomials in $A$ with degree
$(k-l-1)$
and can be expressed as Legendre polynomials,
\be
L^l_k(A)= {2l+1 \over l+1}t^{k-l-1}P_{k-l-1}({2l+3 \over l+2}A) \;\;.
\ee
After rearranging the double summation and some trivial
algebra, we get the following formula for the variation of
the fermionic action:
\ba
\delta S_f&=&
Tr(Z_1 \otimes X_2^{\dagger} \gamma_5 \delta H_W)
+Tr(Z_2 \otimes X_1^{\dagger} \gamma_5 \delta H_W)
\nonumber \\
&-&{1 \over 2c} \sum^{N_{cut}}_{l=0} ({2l+1 \over l+1})Tr[(H_Wx_{2,l} \otimes 
T^{\dagger}_{1,l})\gamma_5\delta H^{\dagger}_W]
\nonumber \\
&-&{1 \over 2c} \sum^{N_{cut}}_{l=0} ({ 2l+1 \over l+1})Tr[x_{2,l} \otimes 
(H_W\gamma_5T_{1,l})^{\dagger}\delta H_W] 
\nonumber \\
&-&{1 \over 2c} \sum^{N_{cut}}_{l=0} ({2l+1 \over l+1})Tr[(H_Wx_{1,l} \otimes 
T^{\dagger}_{2,l})\gamma_5\delta H^{\dagger}_W]
\nonumber \\
&-&{1 \over 2c} \sum^{N_{cut}}_{l=0} ({ 2l+1 \over l+1})Tr[x_{1,l} \otimes 
(H_W\gamma_5T_{2,l})^{\dagger}\delta H_W] \;\;,
\nonumber \\
Z_i&=& (H^{\dagger}_WH_W)^{-1/2}X_i \;\;,\;\;\;\;
x_{i,l}= t^l(P_l(A))X_i \;\;,
\nonumber \\
T_{i,l}&=& \sum^{N-l-1}_{m=0} {\cal S}^{(m)}_{i,l} \;\;,\;\;\;\;
{\cal S}^{(m)}_{i,l} =
t^m P_{m}(({2l+3 \over l+2})\gamma_5A\gamma_5)H_W X_i \;\;.
\ea
Therefore, the following expression for the fermionic force is obtained:
\ba
F_{\mu}(x)&=&
tr_{Dirac}(Z_1(x+\mu) \otimes X_2^{\dagger}(x) \gamma_5(1-\gamma_{\mu})
+(Z_2(x+\mu) \otimes X_1^{\dagger}(x) \gamma_5(1-\gamma_{\mu}))
\nonumber \\
&-&{1 \over 2c} \sum^N_{l=0} ({ 2l+1 \over l+1}) tr_{Dirac}[(H_Wx_{2,l}(x+\mu) 
\otimes T_{1,l}(x))\gamma_5(1+\gamma_{\mu})]
\nonumber \\
&-&{1 \over 2c} \sum^N_{l=0} ({ 2l+1 \over l+1}) tr_{Dirac}[(x_{2,l}(x+\mu) 
\otimes [H^{\dagger}_WT_{1,l}]^{\dagger}(x))\gamma_5(1-\gamma_{\mu})] 
\nonumber \\
&-&{1 \over 2c} \sum^N_{l=0} ({ 2l+1 \over l+1}) tr_{Dirac}[(H_Wx_{1,l}(x+\mu) 
\otimes T_{2,l}(x))\gamma_5(1+\gamma_{\mu})]
\nonumber \\
&-&{1 \over 2c} \sum^N_{l=0} ({ 2l+1 \over l+1}) tr_{Dirac}[(x_{1,l}(x+\mu) 
\otimes [H^{\dagger}_WT_{2,l}]^{\dagger}(x))\gamma_5(1-\gamma_{\mu})] 
\;\;.
\ea
Since the CPU cost of a simulation program with dynamical
fermions is dominated by the fermion matrix times vector operations.
It becomes clear that 
the above formula for the fermionic force is not very useful
from a practical point of view.
The most CPU consuming part is the calculation of
the vectors $T_{i,l}(x)$, for all values of $l$,
each requiring an iterative procedure, i.e. the calculation
of the quantities ${\cal S}^{(m)}_{i,l}$. This implies that, in order to
calculate the fermionic force, 
the multiplication of the matrix $H^{\dagger}_WH_W$ 
has to be performed $O(N^2_{cut})$ times, where $N_{cut}$ 
can become large.  This would make the calculation of 
the fermionic force too costly.

However, there is a way around this difficulty. 
The price to pay will be some extra
storage. Note that Legendre polynomials $P_m(\beta(l)z)$, $\beta(l)$
being the constant $(2l+1)/(l+1)$, could be
expressed as a linear combination of Legendre polynomials 
of lower or equal degrees  with argument changed to $z$, i.e.
\be
P_m(\beta z)= \sum^m_{j=0} \sigma_{m,j}(\beta) P_j(z) \;\;.
\ee
With this, we  could express the quantities $T_{i,l}$ in the following way:
\ba
T_{i,l}&=&\sum^{N-l-1}_{m=0} f_m(l,t)t^mP_m(\gamma_5A\gamma_5)H_WX_i \;\;,
\nonumber \\
f_m(t,l)&=&\sum^{N-l-1}_{j=m} t^{j-m} \sigma_{j,m}(\beta(l)) \;\;.
\ea
The functions $f_m(t,l)$ are just c-numbers and can
be calculated at the beginning of the simulation.
Therefore,after the vectors $X_i$ are obtained,
one can calculate the vectors $P_m(\gamma_5A\gamma_5)HX_i$ 
once for all values of $m$
and store the resulting vectors. 
Thus, $T_{i,l}$ could be obtained easily without
further iteration of matrix multiplications.
The coefficients $\sigma_{m,j}(l)$ satisfy the following recursion relation:
\be
\sigma_{m,j}=({2m-1 \over m})\beta(l)[({j \over 2j-1})\sigma_{m-1,j-1}
+({j+1 \over 2j+3})\sigma_{m-1,j+1}]-({m-1 \over m}) \sigma_{m-2,j} \;\;,
\ee
where the subscripts $m$ $j$ should satisfy $0\le j \le m$ and 
a zero value is understood whenever an out-of-range subscript is
encountered. Together with $\sigma_{0,0} \equiv 1$, the above recursion
relation completely determines all coefficients $\sigma_{m,j}(\beta(l))$
and therefore the function $f_m(t,l)$.
Now the calculation of the quantity $T_{i,l}$ only requires a 
linear combination of vectors which costs little CPU time.

\subsection{comparison of the two methods}

We now compare the CPU time consumption and memory consumptions of the
two methods discussed so far for the simulation of dynamical 
Ginsparg-Wilson fermions. As is well known, the CPU time 
consumption is proportional to the  number of operations
of the matrix $H_W$ on vectors. 
For each molecular dynamics step in the Hybrid Monte Carlo,
ORAM requires $2N_{CG}(2N_{cg}+1)$ number of matrix multiplications
to obtain the solution vector $X_1$ and $4N_{cg}+4N_r$ more matrix
multiplications to obtain the fermionic force. Here,
$N_{CG}$ is the number of conjugate gradient iterations
needed to obtain the solution $X_1$ and $N_{cg}$ is the
number of conjugate gradient iterations needed to obtain
the vector $(H^{\dagger}_WH_W)+q_{min})^{-1}X_i$ for the
smallest shift $q_{min}$.
Parameter $N_r$ is the order of the polynomials in the optimal
rational approximation.
ORAM also requires to store $N_r$ psudofermion field vectors.
As a comparison, 
FIM requires $2N_{CG}(2N_{cut}+1)$ number of matrix multiplications
to obtain the solution vector $X_1$ and $12N_{cut}$ more matrix
multiplications to obtain the fermionic force, where $N_{cut}$
is the highest order of Legendre polynomials kept in the series
expansion.  FIM also needs to store $2N_{cut}$ psudofermion field vectors.

Concerning the CPU time consumption, both method are more costly
than dynamical simulations with Wilson fermions by a factor
of $2N+1$, where $N=N_{cg}$ for ORAM and $N=N_{cut}$ for FIM.
From the theoretical upper bound of the error, the two methods
behave in a similar manner, $N_{cg} \sim N_{cut}$. Practically,
however, according
to the experience in \cite{edwards}, $N_{cg}$ is usually less
than $N_{cut}$ because the theoretical bound is saturated for
FIM  while it is not for ORAM. Therefore, ORAM is more favorable
compared with FIM when doing simulations 
with dynamical Ginsparg-Wilson
fermions. Concerning the memory consumption, $N_r$ is usually much
less than $N_{cut}$ which would 
again put ORAM in a more favorable place.

It is clear from the above discussion that, if one would like
to accelerate the simulation with dynamical Ginsparg-Wilson fermions,
one has to find ways to decrease $N_{cg}$ in ORAM or $N_{cut}$ in FIM.
These two parameters are mainly determined by the condition number
of the matrix $H^{\dagger}_WH_W$. 
Any preconditioning methods that would decrease the condition
number of the matrix (while still maintaining the
shifted nature of the matrix in ORAM)
 will bring an improvement to the simulation of
dynamical Ginsparg-Wilson fermions.
It should be pointed out that other improvements, for example using 
better integration schemes, would apply to both methods.
Test runs on small lattices are now under 
investigation \cite{chuan2} where these 
algorithmic issues will be further studied.

\section{Conclusions}

In this paper, possibilities of simulating dynamical Ginsparg-Wilson fermions
are discussed.  The formula for
the fermionic force is derived for two specific
implementations of the algorithm, the optimal rational
approximation method (ORAM) and fractional inverter method (FIM).
It turns out that, in both methods, 
simulating dynamical Ginsparg-Wilson fermions are more costly than
simulating dynamical Wilson fermions. The extra CPU time
consumption mainly comes from the
the fractional inversion of the matrix.
In quenched simulations, it has been reported \cite{edwards} 
that ORAM performs better than FIM. In dynamical simulations,
this conclusion remains, both on CPU time consumption and 
memory consumption.
It should be emphasized that, though being more costly, 
the advantage of simulating dynamical Ginsparg-Wilson
fermions over dynamical Wilson fermions or
quenched Ginsparg-Wilson fermions would be a much better
behavior towards the chiral limit.
The feasibility of such simulation has been demonstrated in
this paper using a Hybrid Monte Carlo algorithm.

\section*{Acknowledgments}
This work is supported by the Chinese National Science Foundation 
and by the Natural Science Fund from Peking University.

\end{document}